\documentclass[aps,preprint,floatfix,nofootinbib]{revtex4}
\usepackage{epsfig}
\usepackage{graphicx}

\def\lsim{\mathrel {\vcenter {\baselineskip 0pt \kern 0pt
    \hbox{$<$} \kern 0pt \hbox{$\sim$} }}}
\def\gsim{\mathrel {\vcenter {\baselineskip 0pt \kern 0pt
    \hbox{$>$} \kern 0pt \hbox{$\sim$} }}}
\def\slashchar#1{\setbox0=\hbox{$#1$}           
 \dimen0=\wd0                                 
  \setbox1=\hbox{/} \dimen1=\wd1               
\ifdim\dimen0>\dimen1                        
  \rlap{\hbox to \dimen0{\hfil/\hfil}}      
  #1                                        
  \else                                        
 \rlap{\hbox to \dimen1{\hfil$#1$\hfil}}   
   /                                         
  \fi}                                         %
\def\cpto{\mathrel {\vcenter {\baselineskip 0pt \kern 0pt
    \hbox{$CP$} \kern 0pt \hbox{$\longrightarrow$} }}}
\def\cptof{\mathrel {\vcenter {\baselineskip 0pt \kern 0pt
    \hbox{$~CP$} \kern 0pt \hbox{$\longleftrightarrow$} }}}

\begin{document}

\baselineskip=15pt

\preprint{hep-ph/0512127}
\preprint{OKHEP-05-02}

\title{New $CP$-odd Observable in $H\to t\bar{t}$}

\author{G. Valencia$^a$,\ Yili Wang$^{a,b}$}

\email{valencia@iastate.edu}
\email[]{yiliwa@iastate.edu}

\affiliation{$^{a}$Department of Physics, Iowa State University, Ames, IA 50011, $^{b}$Department of Physics, University of Oklahoma, Norman, OK 73019.}

\date{\today}

\vskip 1cm
\begin{abstract}

We study a $CP$-violating triple product correlation that occurs in the decay of a neutral Higgs boson into $t\bar{t}$ pairs when the Higgs boson does not have a definite $CP$ nature. We consider the 
$H \to t\bar{t}$ decay channel as well as the gluon fusion process $gg\to H \to t\bar{t}$.  The asymmetry in Higgs decay, normalized to the $H \to t\bar{t}$ width, can reach the 6\% percent level. In the gluon fusion process the corresponding normalized asymmetry is smaller by an order of magnitude.  We present a crude estimate of this observable at the LHC.

\end{abstract}

\pacs{PACS numbers: }

\maketitle

\section{Introduction}

In models with more than one Higgs doublet it is possible to have 
CP violation in the coupling of neutral Higgs bosons to fermions. Such CP violation manifests itself in the form of a physical neutral Higgs that has both scalar and pseudo-scalar couplings. The couplings to the top-quark would then be given by the Lagrangian
\begin{eqnarray}
{\cal L}&=&-\frac{m_t}{v}H\bar{t} (A+iB\gamma_5)t, 
\label{tlag}
\end{eqnarray}
where $A,B$ are real and $A=1,~B=0$ corresponds to the standard model with one Higgs doublet. Exploring the possible existence of these couplings will be an important task for the LHC.

The determination of the $CP$ nature of a neutral Higgs boson at the LHC has been studied before in the literature. Gunion and He \cite{Gunion:1996xu} considered the production of a light Higgs boson in association with a $t\bar{t}$ pair finding that the spin-averaged cross-section does not contain terms proportional to $AB$. These terms are also absent in the spin-averaged cross-section for the process $gg\to t\bar{t}$. However, the weak decay of the top quark makes it  possible to study spin correlations because it analyzes the top polarization \cite{Bigi:1980az}. The spin correlations manifest themselves in the angular distributions of the decay products of the $t$ and $\bar{t}$.  These top-quark spin correlations have been the subject of much attention  \cite{spinco}, and in our case they provide an avenue to explore the product of couplings $AB$.

A particular example of a $CP$ violating observable that is proportional to $AB$ is the $T$-odd triple product correlation \cite{Donoghue:1986nn,Donoghue:1987ax} that arises for the decay chain 
$H\to t\bar{t}\to b\bar{b}W^+W^-$:
\begin{equation}
\epsilon(p_t,p_{\bar{t}},p_b,p_{\bar{b}}) =
\epsilon(p_{W^+},p_{W^-},p_b,p_{\bar{b}})
= \sqrt{s} \vec{p}_{\bar{b}}\cdot 
(\vec{p}_t\times\vec{p}_b)
\label{cpobs}
\end{equation}
where the last equality follows in the $t\bar{t}$ (parton) center of mass frame. In this frame, the different momenta transform under $CP$ as,
\begin{eqnarray}
\vec{p}_t \ \cpto \ 
-\vec{p}_{\bar{t}}\ =\ \vec{p}_t \ , \ \ 
\vec{p}_b \ \cptof \ -\vec{p}_{\bar{b}}.
\end{eqnarray}
From these transformation properties it follows that this triple product correlation is a $CP$-odd observable:
\begin{eqnarray}
\vec{p}_t\cdot 
(\vec{p}_b\times\vec{p}_{\bar{b}})& \cpto & -\vec{p}_t\cdot 
(\vec{p}_b\times\vec{p}_{\bar{b}}).
\end{eqnarray}
The correlation can then be used to signal and measure $CP$ violation in Higgs decay as well as in the gluon fusion process 
$gg\to t\bar{t}\to b\bar{b}W^+W^-$. 
At the LHC the $pp$ initial state is not a $CP$ eigenstate and the signal could receive contributions from $CP$ conserving backgrounds.  
This has been discussed previously for different $CP$-odd observables considered in the literature~\cite{other}. In particular, the $CP$ odd correlation involving the $t\bar{t}$ spin and momenta that underlies our asymmetry has been previously discussed by Bernreuther and Brandenburg \cite{bern}. 

In this paper we calculate the $CP$ violating observable Eq.~\ref{cpobs} for  both the Higgs decay and gluon fusion processes. We also present a crude estimate for the LHC but leave the details of a more realistic simulation, including the study of background, to a future publication.

\section{Higgs decay}

We begin by considering the process $H\to t \bar{t}\to b\bar{b}W^+W^-$ for the interaction given by Eq.~\ref{tlag}. 
The total Higgs decay width into a $t\bar{t}$ pair is then given by
\begin{equation}
\Gamma(H\to t\bar{t})=N_c\frac{M_H}{32\pi}\frac{g^2m_t^2}{M_W^2}
\sqrt{1-\frac{4m_t^2}{M_H^2}}
\left[\left(|A|^2+|B|^2\right)\left(1-\frac{2m_t^2}{M_H^2}\right)-2
\left(|A|^2-|B|^2\right)\frac{m_t^2}{M_H^2}\right].
\end{equation}

The decay distribution for this process contains a triple product correlation of the form Eq.~\ref{cpobs}. This CP-violating correlation can be obtained in a  straightforward manner; after summing over the spin of the $b$ and $\bar{b}$ as well as the $W^\pm$ polarizations (collectively denoted by $s$) and the quark color, we find:
\begin{eqnarray}
\sum_{s}\left|{\cal M}\right|^2_{tpc} = N_c g^6\frac{m_t^8}{2M_W^6}\left(1-2\frac{M_W^2}{m_t^2}\right)^2 AB \left(\frac{\pi}{m_t\Gamma_t}\right)^2
\epsilon(p_t,p_{\bar{t}},p_b,p_{\bar{b}}) 
\delta(p_t^2-m_t^2)\delta(p_{\bar{t}}-m_t^2)
\label{hmatcp}
\end{eqnarray}
For the intermediate $t\bar{t}$ states we have used the narrow width approximation.  We assume that the $t$-decay occurs as in the standard model and that $V_{tb}=1$, therefore 
\begin{equation}
\Gamma_t = \frac{g^2m_t^3}{64\pi M_W^2}
\left(1-\frac{M_W^2}{m_t^2}\right)^2\left(1+2\frac{M_W^2}{m_t^2}\right).
\end{equation}

To proceed 
we write the four body phase space choosing as independent variables the $t$ angles in the Higgs rest frame, the $b$ angles in the $t$ rest frame and the $\bar{b}$ angles in the $\bar{t}$ rest frame.  The integration range for this choice of angles is not constrained and we write 
\begin{equation}
d\Gamma = \frac{\sum_s|{\cal M}|^2}{(2\pi)^8}\frac{1}{2M_H}
\frac{|\vec{p}_t|d\Omega_t}{4M_H}
\frac{|\vec{p}_b|^\star d\Omega^\star_b}{4m_t}\frac{|\vec{p}_{\bar{b}}|^{\star\star} d\Omega^{\star\star}_{\bar{b}}}{4m_t} dp_t^2 d^2p_{\bar{t}}
\end{equation}
where $^\star$ denotes expressions evaluated in the rest frame of $t$ and $^{\star\star}$ expressions evaluated in the rest frame of $\bar{t}$.

\begin{figure}[htb]
\includegraphics[width=5 in]{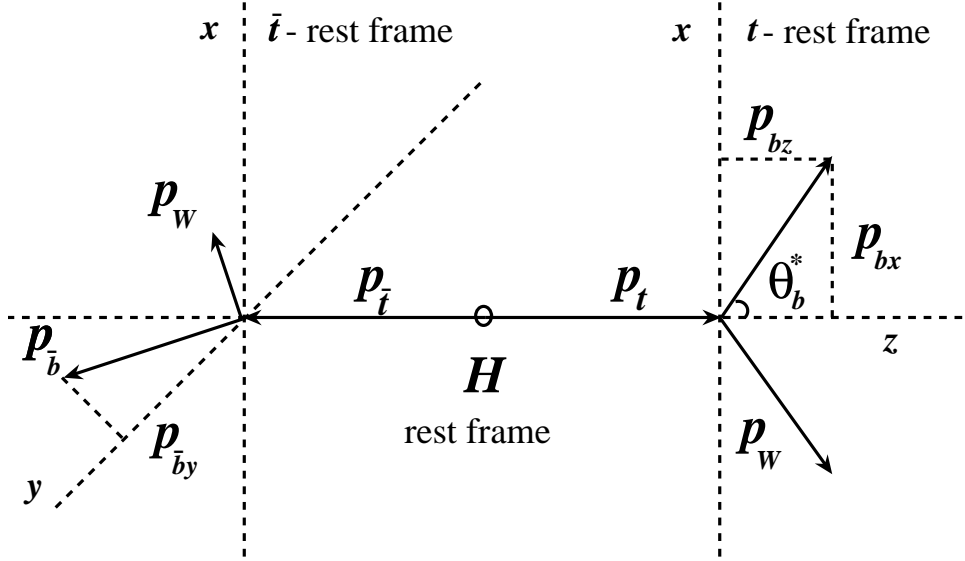}
\caption{Kinematics for the reaction $H\to t\bar{t}\to b\bar{b}W^+W^-$. The $t$-quark momentum, $\vec{p}_t$, defines the $z$-axis. The $t$-quark decay defines the $x-z$ plane and the $\bar{t}$ decay plane forms an angle $\phi^{\star\star}_{\bar{b}}$ with respect to the $t$ decay plane.}
\label{fig:axis}
\end{figure}

In the Higgs rest frame the correlation can be simplified further as 
illustrated in Figure~\ref{fig:axis}.  We use the $t$-quark momentum in this frame to define the $z$-axis and the $t\to b W^+$ decay plane to define the $x-z$ plane.  The $\bar{t}\to\bar{b} W^-$ decay plane then forms an angle $\phi^{\star\star}_{\bar{b}}$ with respect to the $t$-decay plane. 
With these choices we find 
\begin{eqnarray}
\epsilon(p_t,p_{\bar{t}},p_b,p_{\bar{b}}) &=& M_H\vec{p}_{\bar{b}}\cdot (\vec{p}_b\times \vec{p}_t) \nonumber \\
&=& -M_H|\vec{p}_t| p_{bx}p_{\bar{b}y}.
\label{sthr}
\end{eqnarray}
We then note that $p_{bx}$ and $p_{\bar{b}y}$ are invariant under a boost to the $t$ and $\bar{t}$ rest-frames respectively, and use this to calculate them in those frames. This leaves us with a correlation of the form 
\begin{equation}
\epsilon(p_t,p_{\bar{t}},p_b,p_{\bar{b}})= M_H 
\left(\frac{M_H}{2}\sqrt{1-\frac{4m_t^2}{M_H^2}}\right)
\left(\frac{m_t^2-M_W^2}{2m_t}\right)^2\sin\theta^\star_b\sin\theta^{\star\star}_{\bar{b}}\sin\phi^{\star\star}_{\bar{b}}.
\end{equation}
After this, the integration over phase space is straight-forward.  Defining an integrated $CP$ odd observable $\hat{\Gamma}$: 
\begin{eqnarray}
\hat{\Gamma} &\equiv& \int d\Gamma(H\to t \bar{t}\to b\bar{b}W^+W^-) 
~sign(\epsilon(p_t,p_{\bar{t}},p_b,p_{\bar{b}}))
\nonumber \\
&=&\int d\Gamma(H\to t \bar{t}\to b\bar{b}W^+W^-) ~sign(\sin\phi^{\star\star}_{\bar{b}})
\end{eqnarray}
we arrive at our normalized $CP$ asymmetry,
\begin{equation}
A_{CP}\equiv 
\frac{\hat{\Gamma}}{\Gamma} = \frac{\pi}{4}\sqrt{1-\frac{4m_t^2}{M_H^2}}\frac{AB}{|A|^2+|B|^2}\frac{\left(1-\frac{2M_W^2}{m_t^2}\right)^2}{\left(1+\frac{2M_W^2}{m_t^2}\right)^2}\left(\frac{1}
{1-\frac{2m_t^2}{M_H^2}-2\frac{|A|^2-|B|^2}{|A|^2+|B|^2}\frac{m_t^2}{M_H^2}}\right)
\label{higgsob}
\end{equation}
This normalized observable corresponds to the simple counting asymmetry
\begin{equation}
A_{CP}\equiv \frac{N_{events}(\vec{p}_{\bar{b}}\cdot (\vec{p}_b\times\vec{p}_t) > 0) - N_{events}(\vec{p}_{\bar{b}}\cdot (\vec{p}_b\times\vec{p}_t) < 0)}{N_{events}(\vec{p}_{\bar{b}}\cdot (\vec{p}_b\times\vec{p}_t) > 0) + N_{events}(\vec{p}_{\bar{b}}\cdot (\vec{p}_b\times\vec{p}_t) < 0)}
\label{countas}
\end{equation}
in the Higgs rest frame.

Weinberg has shown that there are unitarity constraints on the size of the CP violating couplings of the type of Eq.~\ref{tlag}. For example, in the models discussed by Weinberg~\cite{Weinberg:1990me}, the product $AB$ from Eq.~\ref{tlag} corresponds to what he calls ${\rm Im}~ Z_2$. Assuming that the lightest neutral Higgs eigenstate dominates and that the different scalar vacuum expectation values are comparable, Eqs. 49 or 52 of Ref.~\cite{Weinberg:1990me} then imply,    
\begin{equation}
|AB| \leq \frac{1}{\sqrt{2}}.
\label{unitarity}
\end{equation}
Ref.~\cite{Weinberg:1990me} also shows how to construct models that reach this bound \footnote{For comparison, the first paper in Ref.~\cite{other}, also uses Eq.~\ref{unitarity} and the papers in Ref.~\cite{bern} use the equivalent of $AB \sim  1$}. 
For our numerical estimates of the CP-odd signal, $A_{CP}$, we will choose $A=B=1/\sqrt{2}$ in accordance with Eq.~\ref{unitarity}.

\begin{figure}[!htb]
\begin{center}
\includegraphics[width=6in]{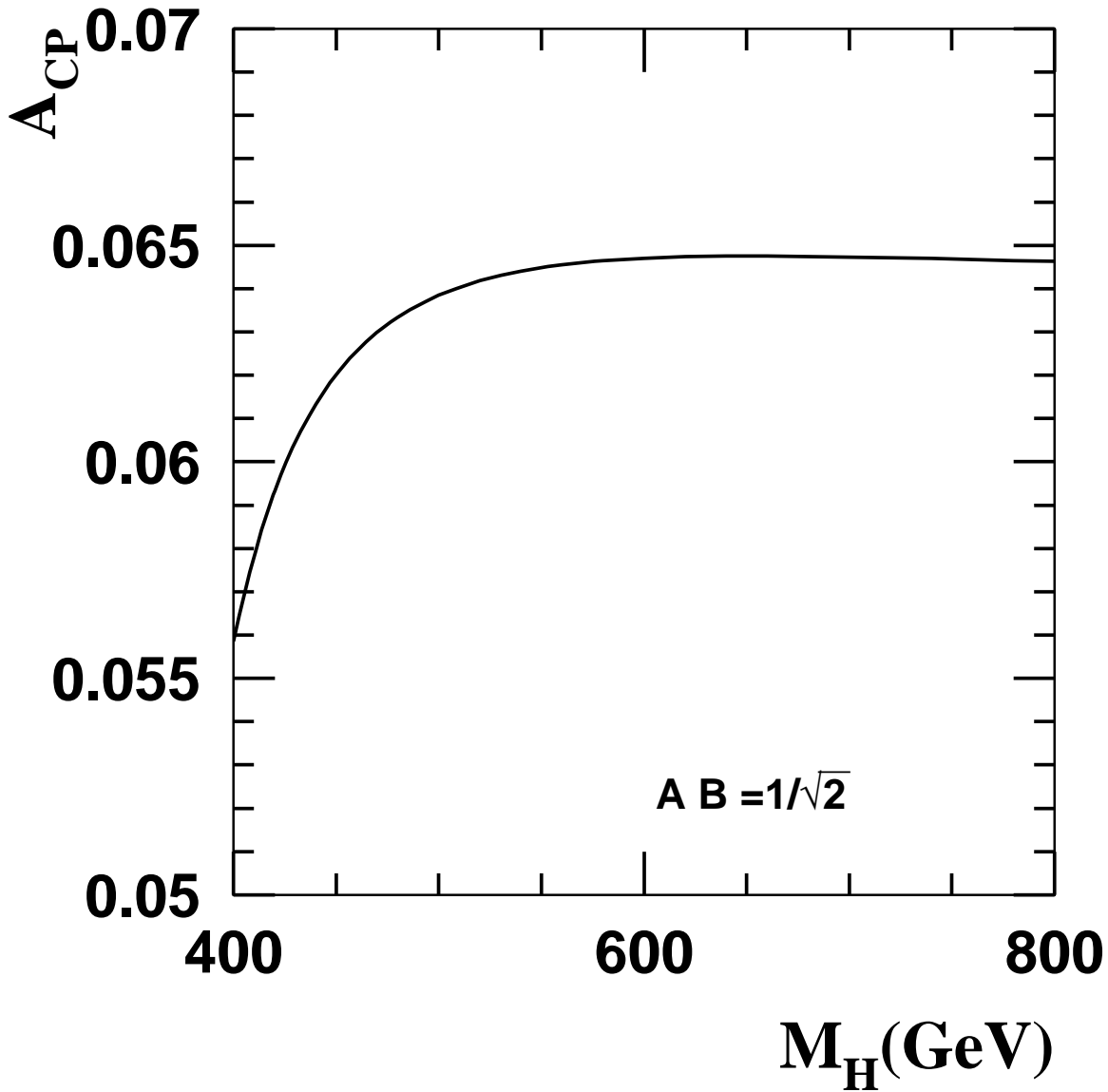}
\end{center}
\caption{Normalized $CP$ violating asymmetry in Higgs decay as given by Eq.~\ref{higgsob}. We use $A=B=1/\sqrt{2}$ from Eq.~\ref{unitarity}.}

\label{fig:higgs}
\end{figure}

Numerically this result is shown in 
Figure~\ref{fig:higgs},
The asymmetry can have either sign depending on the relative sign of $A$ and $B$. It could also be quite large at over $6\%$, and is nearly independent of the Higgs mass (as long as the Higgs is sufficiently heavy to decay into $t\bar{t}$).

\section{$CP$-odd asymmetry in gluon fusion}

\begin{figure}[htb]
\hspace{0.5in}{\includegraphics[width=3 in]{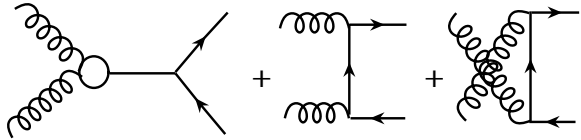}}
\caption{Three diagrams responsible for $CP$ asymmetry in top-quark pair production}
\label{fig:prod}
\end{figure}

In hadron colliders the Higgs boson is produced by gluon fusion through an intermediate top-quark loop. This leads us to consider 
the combined Higgs production and decay through the reaction $gg\to  t\bar{t}$. For this source of $t\bar{t}$ pairs there are  
three relevant diagrams as shown in Figure~\ref{fig:prod}. 
There is the resonant $t\bar{t}$ production shown in the first diagram, and there are additional $t$ and $u$ channel diagrams in which the gluons couple directly to the $t\bar{t}$ pairs which can interfere with the Higgs exchange diagram.
The complete $CP$ violating asymmetry in this case receives two 
contributions. The first contribution arises from the square of the first diagram and corresponds to the asymmetry in Higgs decay modified by factors for the Higgs production vertex. The second contribution arises from the interference between the Higgs exchange diagram and the $t$ and $u$ channel diagrams. Note that there is an additional $s$-channel diagram in which a gluon couples to the $t\bar{t}$ pair. However, this diagram does not interfere with the Higgs exchange diagram and does not contribute to the $CP$ asymmetry.

The effective coupling between the Higgs boson and gluons at one-loop can be written in the form
\begin{eqnarray}
{\cal L}&=&-\left[F_a(s)(-q_1\cdot q_2 g_{\mu\nu}+q_{2\mu}q_{1\nu})+F_b(s)
\epsilon(\mu,\nu,q_1,q_2)\right].
\end{eqnarray}
The two form factors $F_a(s)$ and $F_b(s)$ have been obtained previously in the literature. For the kinematic regime in which the Higgs boson is heavier than a $t\bar{t}$ pair they are given by \cite{Rizzo:1979mf,Gunion:1988mf}
\begin{eqnarray}
F_a &=& \frac{g\alpha_s A}{4\pi M_W}\delta_{ab}\tau_t\left[1+(1-\tau_t)f(\tau_t)\right] \nonumber \\
F_b &=& \frac{g\alpha_s B}{4\pi M_W}\delta_{ab}\tau_t f(\tau_t)
\end{eqnarray}
where $\tau_t=4m_t^2/s$ and for $\tau_t <1$
\begin{equation}
f(\tau)=-\frac{1}{4}\left[\log\left(\frac{1+\sqrt{1-\tau}}{1-\sqrt{1-\tau}}\right)-i\pi\right]^2.
\label{loopff}
\end{equation}
This is the case where $\sqrt{s}$ is large enough to produce top-quark pairs, and therefore the form factors are complex. $CP$ odd but naive-$T$ even observables  (some of which are discussed in Ref.~\cite{other})  require a ``strong phase'' and are thus proportional to the absorptive part of these form factors. Our $CP$-odd triple product correlation, on the other hand, does not require a ``strong phase'' and will be proportional to the real part of these form factors \cite{Donoghue:1986nn}.

\begin{figure}[htb]
\hspace{0.5in}{\includegraphics[width=3in]{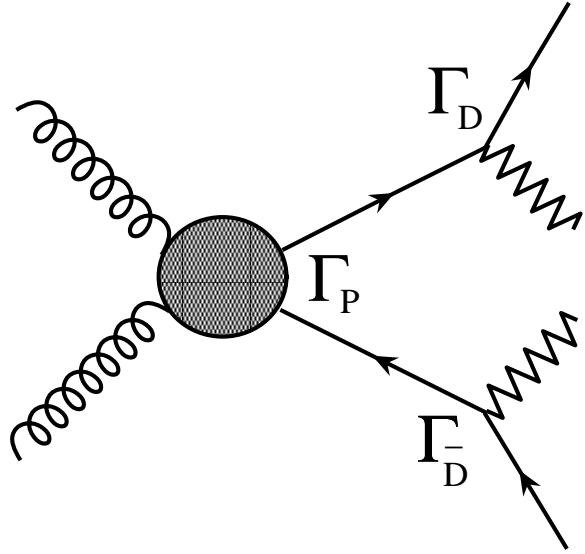}}
\caption{Decomposition of $t\bar{t}$ production and decay vertices with helicity amplitudes.}
\label{fig:fact}
\end{figure}

A convenient way to calculate the $CP$ asymmetry is to consider the process as in Figure~\ref{fig:fact} in the parton CM frame and use helicity amplitudes for the most part. 
The top-quark pair production by the three diagrams in Figure~\ref{fig:prod} is represented by $\Gamma_P$ in Figure~\ref{fig:fact}. The $t$ and $\bar{t}$ decays into $bW$ are represented by $\Gamma_{D,\bar{D}}$. 
It follows that the amplitude can be written as
\begin{equation}
{\cal M} =- \frac{\bar{u}_b\Gamma_D^\mu (\slashchar{p}_t+m_t)\Gamma_P 
(-\slashchar{p}_{\bar{t}}+m_t) \Gamma_{\bar{D}}^\nu v_{\bar{b}}\epsilon^{\star\mu}(W^+)\epsilon^{\star\nu}(W^-)}{(p_t^2-m_t^2)(p_{\bar{t}}^2-m_t^2)}.
\end{equation} 
To construct helicity amplitudes we now replace the numerator of the top-quark (and anti-top-quark) propagator with a sum over polarizations. We work within the narrow width approximation for the $t$ and $\bar{t}$ decays and, therefore, these polarization sums refer to  on-shell $t\bar{t}$ states. As we will see, the $CP$ odd observable arises from the interference of amplitudes in which the intermediate states have different helicities. Our amplitude becomes after this replacement: 
\begin{equation}
{\cal M} = \frac{\bar{u}_b\Gamma_D^\mu (\sum_\lambda u_{t\lambda}\bar{u}_{t\lambda})\Gamma_P (\sum_\sigma v_{\bar{t}\sigma}\bar{v}_{\bar{t}\sigma})
 \Gamma_{\bar{D}}^\nu v_{\bar{b}}\epsilon^{\star\mu}(W^+)\epsilon^{\star\nu}(W^-)}{(p_t^2-m_t^2)(p_{\bar{t}}^2-m_t^2)}.
\end{equation}
We now 
sum over the $W^+$ and the $W^-$ polarization as well as the $b$ and the $\bar{b}$ spin to obtain:
\begin{eqnarray}
\left| {\cal M}\right|^2 &=& \frac{g^4}{16}\left(\frac{\pi}{m_t\Gamma_t}\right)^2\delta(p_t^2-m_t^2)\delta(p_{\bar{t}}-m_t^2)\sum_\lambda \sum_{\lambda'} \sum_\sigma\sum_{\sigma'}\nonumber \\
&\cdot&\bar{u}_{t\lambda'}\gamma_\mu \slashchar{p}_b
\gamma_{\nu}(1-\gamma_5) u_{t\lambda}
\left(-g^{\mu\nu}+\frac{p_{W^+}^\mu p_{W^+}^\nu}{M_W^2}\right)
\nonumber \\
&\cdot& \bar{v}_{\bar{t}\sigma} \gamma_{\mu'} \slashchar{p}_{\bar{b}} \gamma_{\nu'}(1-\gamma_5) v_{\bar{t}\sigma'}
\left(-g^{\mu'\nu'}+\frac{p_{W^-}^{\mu'} p_{W^-}^{\nu'}}{M_W^2}\right)
\nonumber \\
&\cdot&\bar{u}_{t\lambda}\Gamma_Pv_{\bar{t}\sigma}\ \ 
\bar{v}_{\bar{t}\sigma'}\gamma^0\Gamma^\dagger_P\gamma^0 u_{t\lambda'}
\label{msq}
\end{eqnarray}
and we have already used the standard model decay vertex $\Gamma_D^\mu = g\gamma^\mu P_L/\sqrt{2}$. 

The last line in Eq.~\ref{msq}, corresponding to the production vertex,  is more easily computed in the helicity basis and we write it as
\begin{eqnarray}
\bar{u}_{t\lambda} \Gamma_P v_{\bar{t}\sigma}\bar{v}_{\bar{t}\sigma'}\gamma^0\Gamma^\dagger_P \gamma^0 u_{t\lambda'}
\to \frac{1}{4}\sum_{\lambda_1,\lambda_2} {\cal M_P}(\lambda_1,\lambda_2,\lambda,\sigma)
{\cal M_P}^\star(\lambda_1,\lambda_2,\lambda',\sigma')
\end{eqnarray}
where $M_P$ represents the helicity amplitudes for $gg\to t \bar{t}$ 
and the factor $1/4$ arises when we average over the gluon helicities.
These helicity amplitudes are summarized in Table~\ref{tab:ggtt}.

The second line of Eq.~\ref{msq}, corresponding to the $t$-decay,  can be simplified as follows
\begin{eqnarray}
&&\bar{u}_{t\lambda'}\gamma_\mu \slashchar{p}_b
\gamma_{\nu}(1-\gamma_5) u_{t\lambda}
\left(-g^{\mu\nu}+\frac{p_{W^+}^\mu p_{W^+}^\nu}{M_W^2}\right)
\nonumber \\ &\to & 
\bar{u}_{t\lambda'}\left(m_t\left(\frac{m_t^2}{M_W^2}-1\right)+
\slashchar{p}_b\left(2-\frac{m_t^2}{M_W^2}\right)\right)
(1-\gamma_5) u_{t\lambda}.
\label{tdec}
\end{eqnarray}
The $\bar{t}$-decay factor simplifies in a similar manner
\begin{eqnarray}
&&
 \bar{v}_{\bar{t}\sigma} \gamma_{\mu'} \slashchar{p}_{\bar{b}} \gamma_{\nu'}(1-\gamma_5) v_{\bar{t}\sigma'}
\left(-g^{\mu'\nu'}+\frac{p_{W^-}^{\mu'} p_{W^-}^{\nu'}}{M_W^2}\right)
\nonumber \\
&\to& \bar{v}_{\bar{t}\sigma}\left(m_t\left(1-\frac{m_t^2}{M_W^2}\right)+
\slashchar{p}_{\bar{b}}\left(2-\frac{m_t^2}{M_W^2}\right)\right)
(1-\gamma_5) v_{\bar{t}\sigma'}.
\label{tbdec}
\end{eqnarray}
In these last two factors, Eqs.~\ref{tdec},~\ref{tbdec},
only the terms proportional to $\slashchar{p}_b$ and $\slashchar{p}_{\bar{b}}$ contribute to the $CP$-odd asymmetry (as otherwise there are not enough independent four vectors to form the triple product correlation). The surviving terms are also calculated as a function of the $t$ and $\bar{t}$ helicities  and we use the definitions, 
\begin{eqnarray}
{\cal T}_t(\lambda',\lambda) &\equiv & \bar{u}_{t\lambda'}\slashchar{p}_b(1-\gamma_5) u_{t\lambda} \nonumber \\
{\cal T}_{\bar{t}}(\sigma,\sigma') &\equiv & \bar{v}_{\bar{t}\sigma}\slashchar{p}_{\bar{b}}(1-\gamma_5) v_{\bar{t}\sigma'}.
\label{deftfactors}
\end{eqnarray}
These factors are given explicitly in Table~\ref{tab:tbw}. With all this  we arrive at the result,
 \begin{eqnarray}
\left| {\cal M}\right|^2 &=& \frac{g^4}{64}
\left(\frac{\pi}{M_t\Gamma_t}\right)^2 
\left(2-\frac{m_t^2}{M_W^2}\right)^2
\delta(p_t^2-m_t^2)
\delta(p_{\bar{t}}^2-m_t^2) \nonumber \\
&\cdot &\sum_{\lambda_1,\lambda_2,\lambda,\lambda',\sigma,\sigma'} {\cal M_P}(\lambda_1,\lambda_2,\lambda,\sigma)
{\cal M_P}^\star(\lambda_1,\lambda_2,\lambda',\sigma')
{\cal T}_t(\lambda',\lambda) {\cal T}_{\bar{t}}(\sigma,\sigma')
\label{finfor}
\end{eqnarray}

To evaluate this expression we use Table~\ref{tab:ggtt} and Table~\ref{tab:tbw} in the appendix. 
They are calculated in the parton ($gg$) center of mass frame where the $t$ and $b$ four-momenta are parametrized by 
\begin{equation}
p_t = \frac{\sqrt{s}}{2}(1,\beta\sin\theta,0,\beta\cos\theta),\ \ 
p_b = E_b(1,\sin\theta_b\cos\phi_b,\sin\theta_b\sin\phi_b,\cos\theta_b),
\label{bmom}
\end{equation}
and similarly for the $\bar{b}$. $\beta=\sqrt{1-4m_t^2/s}$ is the usual kinematic factor. We emphasize that these angles correspond to the parton CM frame, {\it unlike} the angles used in Section~II, which were given in the $t$ and $\bar{t}$ rest frames. 
In terms of these kinematic variables the desired triple product correlation reads 
\begin{eqnarray}
&&\epsilon(p_{W^+},p_{W^-},p_b,p_{\bar{b}}) = 
\epsilon(p_{t},p_{\bar{t}},p_b,p_{\bar{b}})=-\frac{s}{2}E_b E_{\bar{b}}\sqrt{\left(1-\frac{4m_t^2}{s}\right)}\nonumber \\
&&\left(\sin(\theta)\cos(\theta_{\bar{b}})  \sin(\theta_b) \sin(\phi_b)-
 \cos(\theta) \cos(\phi_{\bar{b}}) \sin(\theta_b) \sin(\theta_{\bar{b}})\sin(\phi_b)\right.\nonumber \\ 
 &&\left . -\sin(\theta)\cos(\theta_b)\sin(\theta_{\bar{b}})\sin(\phi_{\bar{b}})+\cos({\theta})\cos(\phi_b)\sin(\theta_b)\sin(\theta_{\bar{b}})\sin(\phi_{\bar{b}})\right).
 \label{tpexplicit}
\end{eqnarray}

The first contribution to this correlation, from the square of the Higgs exchange diagram, can be extracted from Eq.~\ref{finfor}. Its corresponding color factor is $\delta_{A,B}\delta_{A,B}N_c/8^2=N_c/8$ and we find:
\begin{eqnarray}
\left|{\cal M}\right|^2_{H-tpc} &=& 
\frac{g^2\alpha_s^2}{1024\pi^2M_W^2}\tau_t^2\left(
A^2\left| \left[1+(1-\tau_t)f(\tau_t)\right]\right|^2+B^2\left|f(\tau_t)\right|^2\right)\frac{s^2}{(s-M_H^2)^2+\Gamma_H^2 M_H^2}\nonumber \\
&\cdot & N_c g^6\frac{m_t^8}{2M_W^6}\left(1-2\frac{M_W^2}{m_t^2}\right)^2 AB \left(\frac{\pi}{m_t\Gamma_t}\right)^2 \nonumber \\
&\cdot& \epsilon(p_t,p_{\bar{t}},p_b,p_{\bar{b}}) 
\delta(p_t^2-m_t^2)\delta(p_{\bar{t}}-m_t^2).
\label{hmatcpgg}
\end{eqnarray}
We have written Eq.~\ref{hmatcpgg} explicitly in terms of Eq.~\ref{hmatcp}. This serves as a check of the calculation since this result can also be obtained from multiplying Eq.~\ref{hmatcp} with appropriate factors for the Higgs propagator and the Higgs production vertex. The resulting correlation is not linear in any absorptive phase as expected. The subscript $tpc$ indicates that we have only retained those terms that give rise to the $CP$ violating correlation and dropped all others.

The second contribution to the asymmetry results from the interference of either the $t$-channel or $u$-channel $gg\to t\bar{t}$ diagram with the $s$-channel Higgs diagram. It can also be extracted from Eq.~\ref{finfor} but this time the color factor is equal to
$\delta_{A,B}{\rm Tr}(T^AT^B)/8^2=1/16$. 
The desired correlation is found to be proportional to the coupling constant combination $b\Re[ (F_a)]+a \Re[( F_b)]$, once again no absorptive phase is required to generate the $CP$ violating signal. Our result for this contribution to the asymmetry in the parton center of mass frame is 
\begin{eqnarray}
\left|{\cal M}\right|^2_{int-tpc} &=& -256\pi^4\alpha_s^2 \ AB\frac{g^2}{M_W^2}
\frac{\left(1-\frac{2 M_W^2}{m_t^2}\right)^2}
{\left( 1-\frac{M_W^2}{m_t^2}\right)^4\left(1+\frac{2M_W^2}{m_t^2}\right)^2}\frac{\tau_t\left[1+(2-\tau_t)\Re[f(\tau_t)]\right]}
{\left(s-M_H^2\right)\left(\sin^2\theta+\tau_t\cos^2\theta\right)} 
\nonumber \\ &\cdot&
\epsilon(p_{t},p_{\bar{t}},p_b,p_{\bar{b}}) 
\delta(p_t^2-m_t^2)
\delta(p_{\bar{t}}^2-m_t^2) 
\label{cpinter}
\end{eqnarray}
The complete asymmetry in gluon fusion is the sum of Eq.~\ref{hmatcpgg}, and Eq.~\ref{cpinter}. In the $t\bar{t}$ center of mass frame it corresponds, once again, to the simple counting asymmetry of Eq.~\ref{countas}. Of course, these results need to be folded with appropriate parton distribution functions and evaluated in the lab frame to arrive at physical observables for hadron colliders. We do this for the case of the LHC in the next section.

\section{Results for the LHC}

\begin{figure}[!htb]
\begin{center}
\includegraphics[width=6in]{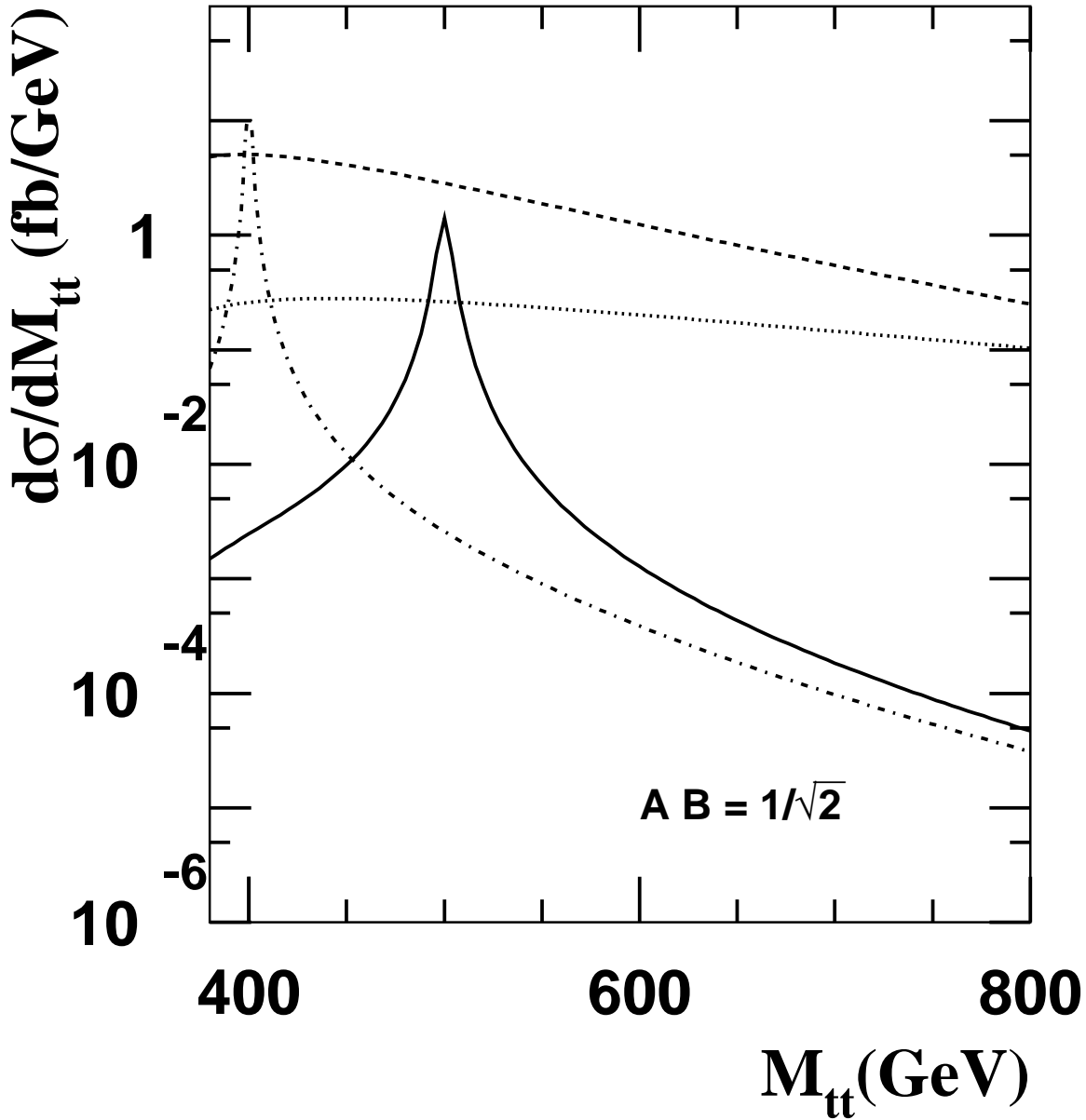}
\end{center}
\caption{$t\bar{t}$ production mechanisms at the LHC. The larger cross-section indicated by the dashed line corresponds to gluon fusion. The dotted line indicates light $q\bar{q}$ annihilation. We also show the resonant Higgs production for two different values of the Higgs mass.}
\label{fig:LHC}
\end{figure}

At the LHC there are three main mechanisms for $t\bar{t}$ production:   gluon fusion, resonant production via an intermediate Higgs boson and  
light $q\bar{q}$ annihilation. We reproduce in Figure~\ref{fig:LHC} the relative magnitudes of these mechanisms with Higgs couplings given by Eq.~\ref{tlag}, with $A=B=1/\sqrt{2}$ per Eq.~\ref{unitarity}, as a function of the $t\bar{t}$ pair invariant mass \footnote{Throughout this section we will use the CTEQ6M parton distribution functions and $\sqrt{S}=14$~TeV. 
}. The resonant production is shown for two values of the Higgs mass. 
This figure illustrates the well known result that the gluon fusion contribution dominates for most of the $m_{tt}$ range (except very close to the Higgs resonance where the two mechanisms are comparable) . The light $q\bar{q}$ production mechanism is about an order of magnitude smaller for $M_{tt} \lsim 750$~GeV.

We first study the $CP$ violating asymmetry generated by the resonant production of a Higgs at the LHC. This contribution to the  total asymmetry is obtained from Eq.~\ref{hmatcpgg} after we use the narrow width approximation for the Higgs propagator. In the lab frame the correlation cannot be rewritten as the simpler triple product $\vec{p}_{\bar{b}}\cdot (\vec{p}_b\times\vec{p}_t)$ of Eq.~\ref{cpobs}. Instead, the physical asymmetry is now defined by 
\begin{equation}
A_{CP}\equiv \frac{N_{events}\left( \epsilon(p_t,p_{\bar{t}},p_b,p_{\bar{b}}) >0 \right)-N_{events}\left( \epsilon(p_t,p_{\bar{t}},p_b,p_{\bar{b}}) <0 \right)}{N_{events}\left( \epsilon(p_t,p_{\bar{t}},p_b,p_{\bar{b}}) >0 \right)+N_{events}\left( \epsilon(p_t,p_{\bar{t}},p_b,p_{\bar{b}}) <0 \right)},
\label{fullacp}
\end{equation}
and requires full reconstruction of four four-momenta.

It is possible to construct a different $CP$-odd correlation that generalizes the triple product of Eq.~\ref{cpobs} to the lab frame 
and that only requires the reconstruction of the directions of the relevant momenta. One such example is 
\begin{equation}
\hat{A}_{CP}\equiv \frac{N_{events}\left(
(\vec{p}_{\bar{t}}-\vec{p}_t)\cdot(\vec{p}_b\times\vec{p}_{\bar{b}})
> 0\right) - N_{events}\left((\vec{p}_{\bar{t}}-\vec{p}_t)\cdot(\vec{p}_b\times\vec{p}_{\bar{b}})< 0\right)}{N_{events}\left(
(\vec{p}_{\bar{t}}-\vec{p}_t)\cdot(\vec{p}_b\times\vec{p}_{\bar{b}})
> 0\right) + N_{events}\left((\vec{p}_{\bar{t}}-\vec{p}_t)\cdot(\vec{p}_b\times\vec{p}_{\bar{b}})< 0\right)}.
\label{trip}
\end{equation}

\begin{figure}[!htb]
\begin{center}
\includegraphics[width=6in]{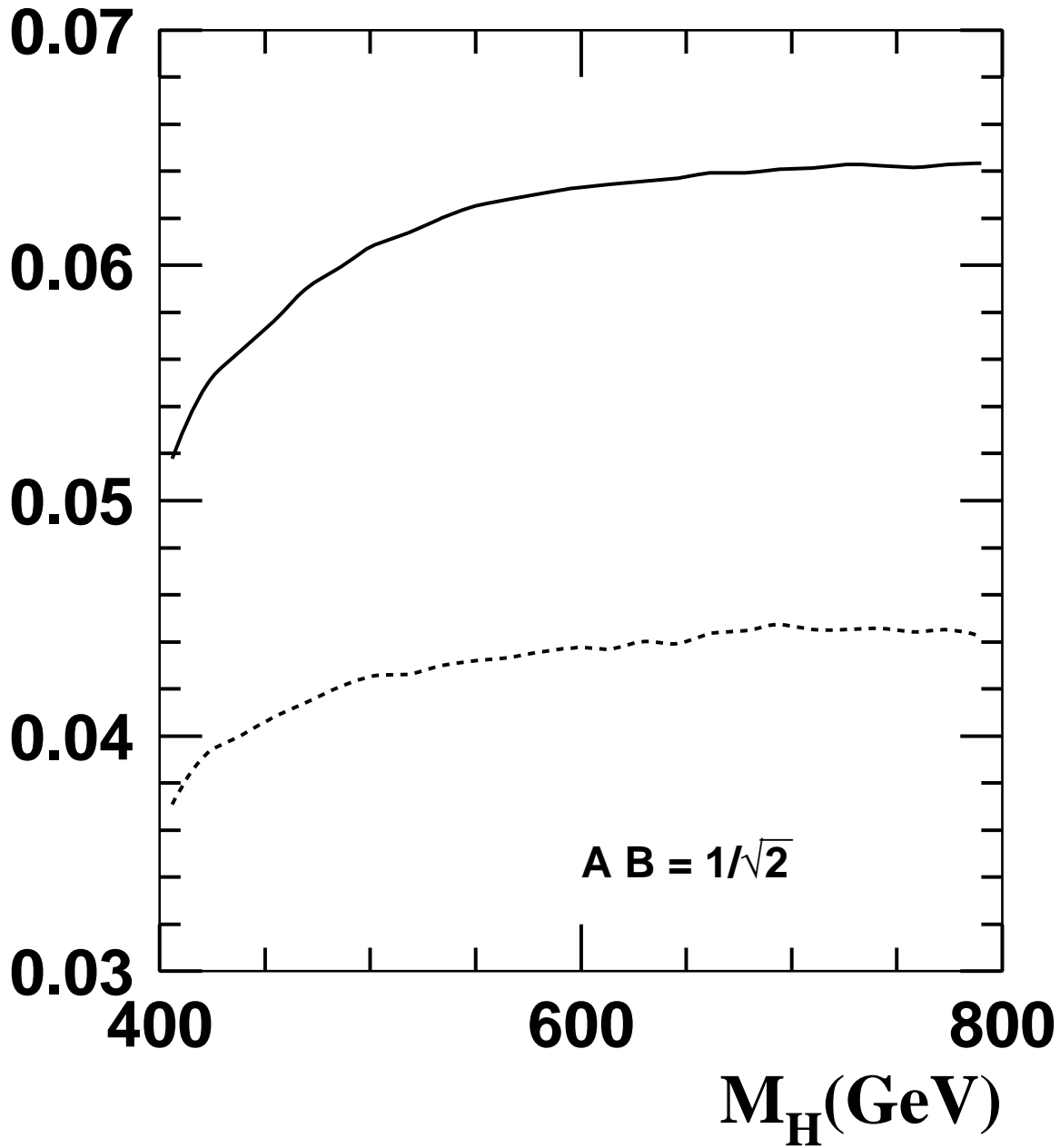}
\end{center}
\caption{$CP$ Asymmetry  for $s$-channel Higgs production and subsequent decay at the LHC. The solid and dotted curves correspond to the asymmetries $A_{CP}$, Eq.~\ref{fullacp}, and $\hat{A}_{CP}$, Eq.~\ref{trip} respectively. }
\label{fig:higgslhc}
\end{figure}

In Figure~\ref{fig:higgslhc} we show both $A_{CP}$ (solid line) and 
$\hat{A}_{CP}$ (dotted line) as a function of the Higgs mass. We see that the production factors in Eq.~\ref{hmatcpgg} cancel out and $A_{CP}$ is the same as calculated for Higgs decay (in the narrow-width s-channel approximation) and shown in Figure~\ref{fig:higgs}. On the other hand, there is a dilution in the asymmetry of about 30\% when using the form $\hat{A}_{CP}$. 

\begin{figure}[!htb]
\begin{center}
\includegraphics[width=6in]{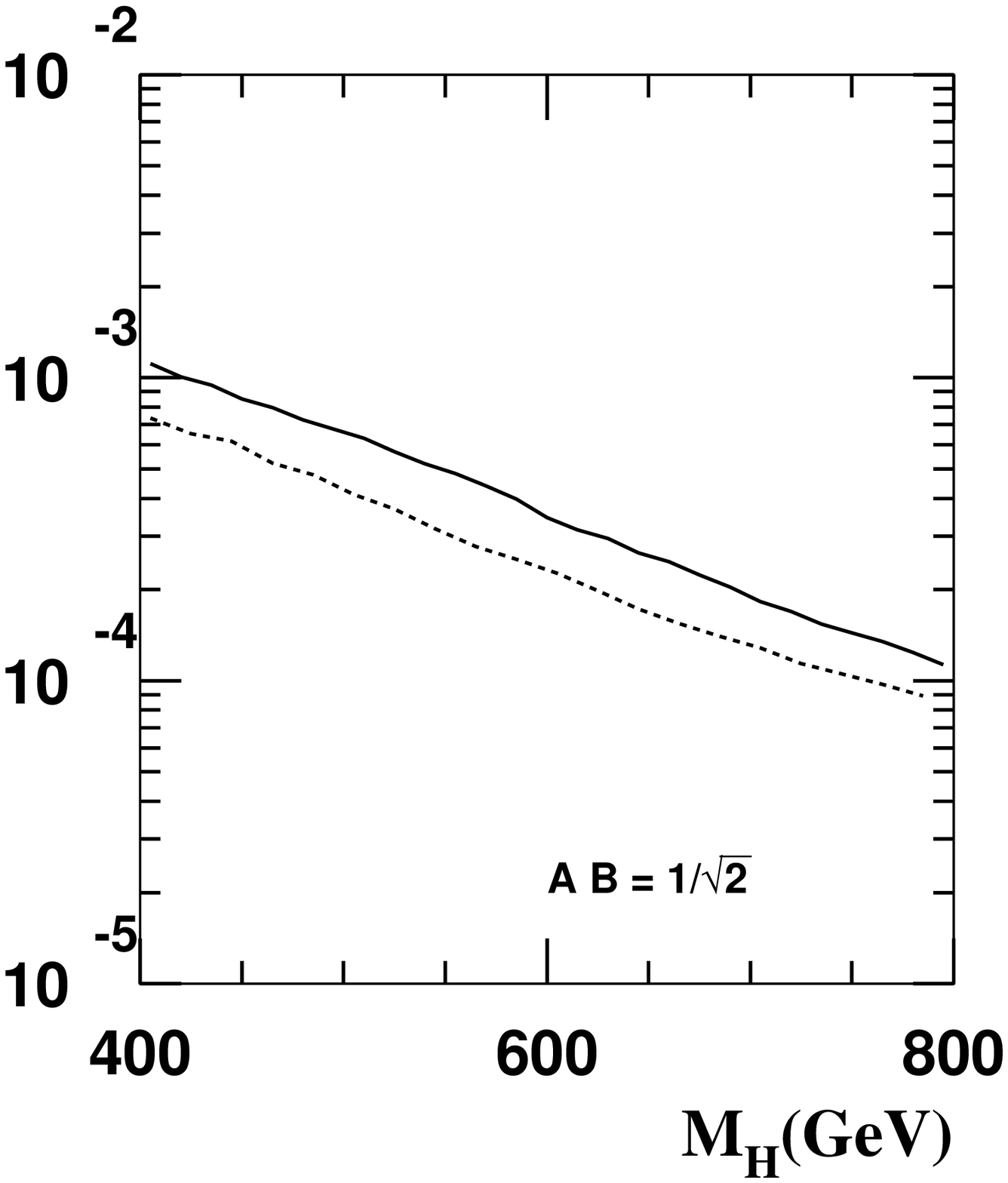}
\end{center}
\caption{Normalized $CP$ violating asymmetry at the LHC for $A,B=1/\sqrt{2}$. The solid and dotted curves correspond to the asymmetries $A_{CP}$, Eq.~\ref{fullacp}, and $\hat{A}_{CP}$, Eq.~\ref{trip}  respectively. In this case both contributions to the asymmetry are included and they are normalized to the total $pp\to t\bar{t}$ cross-section.}
\label{fig:total}
\end{figure}

The complete asymmetry at the LHC is obtained by adding Eqs.\ref{hmatcpgg},~and~\ref{cpinter} (and folding in the corresponding parton distribution functions) \footnote{There is also a small contribution to the asymmetry from light $q\bar{q}$ Higgs production which we neglect in our discussion.}. The normalized asymmetry is shown in Figure~\ref{fig:total}. We present results for $A_{CP}$ (solid line) and $\hat{A}_{CP}$ (dotted line). 
Both of them are significantly smaller than in Higgs decay because the total cross-section for $t\bar{t}$ production (which appears in the denominator of Eq.~\ref{fullacp} and of Eq.~\ref{trip}) is much larger than the Higgs production cross-section.  The asymmetry does not increase as much as the cross-section because the gluon fusion mechanism only contributes to it through  interference with the Higgs  resonance. We find once more that $\hat{A}_{CP}$ is smaller than 
${A}_{CP}$ by about 30\%.
Notice that the light $q\bar{q}\to t\bar{t}$ process can only contribute a very small amount to the asymmetry. This is because it occurs dominantly through one gluon exchange which does not interfere with the resonant Higgs production channel.

\section{Summary and Conclusions}

We have presented a new $CP$ odd observable in the form of a triple product correlation  that is sensitive to a mixture of scalar and pseudoscalar couplings of a Higgs boson to the top-quark. We have evaluated this observable in simple Higgs decay as well as in gluon fusion production of $t\bar{t}$ pairs and illustrated that the resulting asymmetry can occur at the 6\% percent level if the intrinsic $CP$ violation is near its unitarity bound. To reconstruct the asymmetry it is necessary to observe the 
$t$ (or $\bar{t}$) decay plane as well as the direction of the $\bar{b}$ 
(or $b$) momentum. 

The observable originates in a $t\bar{t}$ spin correlation that is analyzed by the weak decay of the top-quark. For this reason it will only apply to very heavy Higgs bosons that can decay into $t\bar{t}$ pairs. For lighter Higgs bosons it is possible to construct a similar observable in the $H\to \tau^+\tau^-$ channel. Although this extension is straightforward, it is not clear that it will be possible to identify $\tau^\pm$ pairs and their decay planes at high energy colliders.

We have estimated the size of the asymmetry in models with multiple Higgs bosons as the source of $CP$ violation. However, the asymmetries $A_{CP}$ from Eq.~\ref{fullacp}, and $\hat{A}_{CP}$ from Eq.~\ref{trip} are more general than this and can be generated by other mechanisms of $CP$ violation. In this sense they constitute a new tool to search for $CP$ violation at future colliders.

We have illustrated a preliminary application of our result to $pp$ collisions at the LHC. In this case reconstruction of the signal $A_{CP}$ in the lab frame is harder and there is a dilution if one uses the simpler signal  $\hat{A}_{CP}$ instead. For leptonic $W$ decays  the direction of the $t (\bar{t})$ cannot be reconstructed. It is then necessary to define a different asymmetry. One possibility is to replace the $t (\bar{t})$ momenta with the momenta of the leptons from the $W$ decay.  Denoting by $\ell^\pm$ an $e^+e^-$ or a $\mu^+\mu^-$ pair one could use, for example,
\begin{equation}
\tilde{A}_{CP}\equiv \frac{
N_{events}\left( (\vec{p}_{\ell^+}-\vec{p}_{\ell^-})\cdot 
(\vec{p}_b\times\vec{p}_{\bar{b}}) > 0\right) - 
N_{events}\left( (\vec{p}_{\ell^+}-\vec{p}_{\ell^-})\cdot
(\vec{p}_b\times\vec{p}_{\bar{b}})< 0\right)}{N_{events}\left(
(\vec{p}_{\ell^+}-\vec{p}_{\ell^-})\cdot(\vec{p}_b\times\vec{p}_{\bar{b}})
> 0\right) + N_{events}\left((\vec{p}_{\ell^+}-\vec{p}_{\ell^-})\cdot(\vec{p}_b\times\vec{p}_{\bar{b}})< 0\right)}.
\label{triptil}
\end{equation}
However, we expect a further dilution of the signal in this case. A further complication at the LHC is that the initial $pp$ state does not transform into itself under $CP$ conjugation and therefore, it is possible to have $CP$ conserving backgrounds to these asymmetries. A discussion of these backgrounds, as well as a more detailed simulation of the asymmetry are necessary for a realistic evaluation of the usefulness of this new observable at the LHC. We leave these considerations to a future publication. There exist several papers in the literature that develop CP asymmetries in similar decay chains to the one considered here in the context of $e^+e^-$ or $\gamma\gamma$ colliders \cite{epem}. The papers of Ref.~\cite{bern} 
\footnote{We thank W. Bernreuther for bringing this work to our attention.} first  discussed the underlying correlation involving the spin and momenta of the $t\bar{t}$ pair. They developed the asymmetry in a slightly different manner that involves the momenta of the secondary leptons. It is difficult to compare their results with our form for the asymmetry without a more detailed numerical simulation. 

\begin{acknowledgments}

This work was supported
in part by the U.S. Department of Energy under contract number DE-FG02-01ER41155. 
The work of Y. W. was also supported in part by the U.S. Department of Energy under Grants No.~DE-FG02-04ER41305 and No.~DE-FG02-03ER46040.
We thank Jian-Wei Qiu for useful conversations.

\end{acknowledgments}

\clearpage

\appendix

\section{Tables of helicity factors}

In this appendix we present our helicity factors. They are calculated in the parton center of mass frame with the first parton ($g$) defining the $z$-axis and $\theta$ being the polar angle of the $t$-quark momentum. These expressions are used in this paper for $pp$ collisions at the LHC, therefore terms that are odd under the interchange $z\leftrightarrow -z$ will not contribute to the differential cross-section after the parton distribution functions are folded in. 

First we present the helicity amplitudes for $gg\to t\bar{t}$. The kinematic factors in Table~\ref{tab:ggtt} refer to the top-quark. Only the first four rows of the table contribute to the asymmetry which requires interference with the $s$-channel Higgs exchange amplitude. The remaining rows are quoted for completeness.
\begin{table}
\caption[Mggtth]{$t+u$ channel $gg\to t\bar{t}$ and $gg\to H\to t\bar{t}$ helicity amplitudes. The former should be multiplied by $g_s^2$ and the latter by $(gm_t/2M_W)$. Color factors have not been included and the kinematic factors refer to the $t$-quark in the parton CM frame:  $\tau_t=4m_t^2/s$, $\beta=\sqrt{1-\tau_t}$, $\gamma=(1-\beta^2)^{-1/2}$ .}
\begin{tabular}{|c|c|c|}\hline
$g,g \rightarrow t \bar{t}$ & $t$+$u$ channels & $s$-channel  \\[1.0ex]\hline
$+,+ \rightarrow +,+$ 
& $\frac{2\sqrt{\tau_t}(1+\beta)}{\beta^2\cos^2\theta-1}$ & $\frac{s^{3/2}}{2(s-M_H^2)}(F_b-iF_a)(B+iA\beta)$
 \\ [1.0ex]
$+,+ \rightarrow -,-$ 
&  $\frac{2\sqrt{\tau_t}(1-\beta)}{\beta^2\cos^2\theta-1}$ & $\frac{s^{3/2}}{2(s-M_H^2)}(F_b-iF_a)(B-iA\beta)$ \\ [1.0ex]
$-,-  \rightarrow +,+$ 
&  $\frac{2\sqrt{\tau_t}(1-\beta)}{1-\beta^2\cos^2\theta}$ 
&  $-\frac{s^{3/2}}{2(s-M_H^2)}(F_b+iF_a)(B+iA\beta)$\\ [1.0ex]
$-,- \rightarrow -,- $
& $\frac{2\sqrt{\tau_t}(1+\beta)}{1-\beta^2\cos^2\theta}$ 
& $-\frac{s^{3/2}}{2(s-M_H^2)}(F_b+iF_a)(B-iA\beta)$\\ [1.0ex]
\hline
$-,+ \rightarrow -,- $ &$\frac{2\beta\sin^2\theta}{\gamma(1-\beta^2\cos^2\theta)}$ & 0 \\ [1.0ex]
$-,+ \rightarrow -,+ $ &$  \frac{2\beta\sin\theta(1+\cos\theta)}{\beta^2\cos^2\theta-1}$ & 0 \\ [1.0ex]
$-,+ \rightarrow +,- $ &$  \frac{2\beta(\cos\theta-1)\sin\theta}{\beta^2\cos^2\theta-1}$ & 0 \\ [1.0ex]
$-,+ \rightarrow +,+ $ &$  \frac{2\beta\sin^2\theta}{\gamma(\beta^2\cos^2\theta-1)}$ & 0 \\ [1.0ex]
$+,- \rightarrow -,- $ &$\frac{2\beta\sin^2\theta}{\gamma(1-\beta^2\cos^2\theta)}$  & 0 \\ [1.0ex]
$+,- \rightarrow -,+ $ &$  \frac{2\beta(\cos\theta-1)\sin\theta}{\beta^2\cos^2\theta-1}$ & 0 \\ [1.0ex]
$+,- \rightarrow +,- $ &$  \frac{2\beta\sin\theta(1+\cos\theta)}{\beta^2\cos^2\theta-1}$ & 0 \\ [1.0ex]
$+,- \rightarrow +,+ $ &$\frac{2\beta\sin^2\theta}{\gamma(\beta^2\cos^2\theta-1)}$ & 0 \\  [1.0ex]
\hline
\end{tabular}
\label{tab:ggtt}
\end{table}

Finally we present in Table~\ref{tab:tbw} the $t$ and $\bar{t}$ decay factors as defined in Eqs.~\ref{deftfactors}. In addition to the top-quark kinematic factors, the results contain the parton center of mass polar and azimuthal angles for the $b$ and $\bar{b}$ momenta as defined in Eq.~\ref{bmom}. Notice that the factors that generate the triple product correlation, $\sin\theta_b\sin\phi_b$ and $\sin\theta_{\bar{b}}\sin\phi_{\bar{b}}$ only occur in the factors with ``mixed'' helicities. That is, the correlation arises from the interference of different helicity amplitudes of the intermediate top-quark. 
\begin{table}
\caption[Mggtt]{$t$ and $\bar{t}$ decay factors as defined in the text. The first column should be multiplied by $E_b$ and the second column by $E_{\bar{b}}$, the $b$ and $\bar{b}$ energies in the parton center of mass frame.}
\begin{tabular}{|c|c|c|}\hline
$\lambda,\lambda'$ & ${\cal T}_t(\lambda,\lambda')$ &  ${\cal T}_{\bar{t}}(\lambda,\lambda')$ \\[1.0ex]\hline
$-,-$ & $ \sqrt{s}(1+\beta)(1-\cos\theta\cos\theta_b -\sin\theta\sin\theta_b\cos\phi_b)$
& $ \sqrt{s}(1-\beta)(1-\cos\theta\cos\theta_{\bar{b}} -\sin\theta\sin\theta_{\bar{b}}\cos\phi_{\bar{b}})$
\\[1.0ex]
$-,+$ & $2m_t (-\sin\theta\cos\theta_b+\cos\theta\sin\theta_b\cos\phi_b-i\sin\theta_b\sin\phi_b)$
& $2m_t (-\sin\theta\cos\theta_{\bar{b}}+\cos\theta\sin\theta_{\bar{b}}\cos\phi_{\bar{b}}+i\sin\theta_{\bar{b}}\sin\phi_{\bar{b}})$
\\[1.0ex]
$+,-$ & $2m_t (-\sin\theta\cos\theta_b+\cos\theta\sin\theta_b\cos\phi_b+i\sin\theta_b\sin\phi_b)$
& $2m_t (-\sin\theta\cos\theta_{\bar{b}}+\cos\theta\sin\theta_{\bar{b}}\cos\phi_{\bar{b}}-i\sin\theta_{\bar{b}}\sin\phi_{\bar{b}})$
\\[1.0ex]
$+,+$ & $\sqrt{s}(1-\beta)(1+\cos\theta\cos\theta_b +\sin\theta\sin\theta_b\cos\phi_b)$
& $ \sqrt{s}(1+\beta)(1+\cos\theta\cos\theta_{\bar{b}} +\sin\theta\sin\theta_{\bar{b}}\cos\phi_{\bar{b}})$
\\[1.0ex]
\hline
\end{tabular}
\label{tab:tbw}
\end{table}

\end{document}